\documentstyle[manuscript,prl,aps]{revtex}
\begin{document}
\title{A Unified Algebraic Approach to Few and Many-Body Hamiltonians having
Linear Spectra}
\author{N. Gurappa, Prasanta K. Panigrahi, and T. Soloman Raju}
\address{School of Physics, 
University of Hyderabad,
Hyderabad,\\ Andhra Pradesh,
500 046 INDIA.}
\maketitle

\begin{abstract} 
We develop an algebraic approach for finding the eigenfunctions of a large
class of few and many-body Hamiltonians, in one and higher dimensions, having
linear spectra. The method presented enables one to exactly map these
interacting Hamiltonians to decoupled oscillators, thereby, giving a precise
correspondence between the oscillator eigenspace and the wavefunctions of
these quantum systems. The symmetries behind the degeneracy structure and the
commuting constants of motion responsible for the quantum integrability of
some of these models are made transparent. Apart from analysing a number of
well-known dynamical systems like planar oscillators with commensurate
frequencies, both with or without singular inverse-square terms and
generalized Calogero-Sutherland type models, we also point out a host of
other examples where the present approach can be profitably employed. We
further study Hamiltonians having Laughlin wavefunction as the ground-state
and establish their equivalence to free oscillators. This reveals the
underlying linear $W_{1+\infty}$ symmetry algebra unambiguously and
establishes the Laughlin wavefunction as the highest weight vector of this
algebra.
\end{abstract}
\draft
\newpage

\begin{center}
{\bf 1. INTRODUCTION}
\end{center}

Few and many-body Hamiltonians having linear energy spectra appear in the
description of various physical phenomena. A few well-known examples, sharing
this common property, are electrons in a constant magnetic field\cite{mag},
isotropic and anisotropic oscillators, with or without singular
inverse-square terms\cite{yu,smo,aojdl,rg,of,roy,luc,qualg} and
Calogero-Sutherland (CS) type systems\cite{calo,suther}. In recent times, the
anisotropic oscillator has attracted considerable attention in the context of
of the descriptions of the intrinsic states of rotating deformed
nuclei\cite{rg,nil}, super-deformed high-spin states of spheroidal nuclear
shapes\cite{ste} and fractional quantum Hall effect\cite{fqhe}. CS model and
its generalizations have also been studied quite extensively in the current
literature. They have found physical applications in various fields such as
the universal conductance fluctuations in mesoscopic systems\cite{meso},
quantum Hall effect\cite{qhe}, wave propagation in stratified
fields\cite{wp}, random matrix theory\cite{suther,rmt}, fractional
statistics\cite{fs}, gravity\cite{gra} and gauge theories\cite{gt}.

The linearity of the energy eigenvalues of these diverse interacting systems
naturally suggests a probable connection with decoupled harmonic oscillators.
This correspondence will, not only provide an elegant explanation for the
linearity of the spectra, but also facilitate the characterization of the
dynamical symmetries underlying these physical systems\cite{ds}. Recently,
two of the present authors have shown that the many-body, Calogero-Sutherland
model (CSM) of both $A_{N-1}$ and $B_N$ types can be mapped into a set of
decoupled harmonic oscillators by means of a series of similarity
transformations\cite{guru1,guru2}. This mapping has led to a straightforward
construction of the eigenfunctions, the commuting constants of motion, and
enables one to write down the decoupled oscillator algebra for the {\it
non-trivially} interacting CSM.

In this paper, we point out that a similar technique can be profitably
employed for mapping a number of dynamical systems with linear spectra, to
decoupled oscillators. As the first example, we establish the exact
correspondence of the much studied, two-dimensional anisotropic oscillator
with commensurate frequencies, both with or without singular inverse-square
terms, with the isotropic one. Making use of this connection, we {\it
unambiguously} show that the symmetry algebra underlying the degeneracies of
the anisotropic oscillator is a regular $SU(2)$ algebra. This result is
subsequently extended to the $N$-dimensional case.

Section 3, deals with few-body Hamiltonians of CSM type. The algebraic
structure, eigenfunctions and quantum integrability of a recently studied
four particle model in one-dimension\cite{ruhl} is studied in detail. We then
point out how the present method can be used to construct more general
interacting models of CSM type.

In Section 4, we analyse various two-dimensional interacting Hamiltonins for
electrons under the influence of both electromagnetic and Chern-Simons gauge
fields.  These models have Laughlin wavefunction as the ground-state and are
also relevant for implementing the Jain picture of the fractional quantum
Hall effect\cite{jain}. We explicitly show the connection of these
interacting models with non-interacting harmonic oscillators. This reveals
the underlying linear $W_{1+\infty}$ symmetry algebra with the Laughlin
wavefunction as the highest weight vector of this algebra.

\begin{center}
{\bf 2. ANISOTROPIC OSCILLATORS}
\end{center}

In the following, we show the precise correspondence of the two-dimensional
anisotropic oscillator, both with or without singular inverse-square terms,
to the isotropic one, establish the connection between their respective
Hilbert spaces and write down the $SU(2)$ symmetry algebra underlying the
degeneracies. There have been earlier attempts at explaining this degeneracy
by $SU(2)$ Lie algebra; however, they were not completely satisfactory
because of ambiguities in the definition of the generators\cite{roy,yu,smo}.
This has led many authors to look for non-linear generalization of the
$SU(2)$ algebra\cite{of,luc}. The mapping to isotropic oscillators naturally
leads to an {\it unambiguous} definition of these generators. Later, this
result is extended to the $N$-dimensional case, where the symmetry algebra is
that of $SU(N)$.

The Hamiltonian of the two-dimensional anisotropic oscillator without
singular inverse-square terms is given by $(\hbar = \omega = m = 1)$
\begin{equation} \label{aniso}
H = - \frac{1}{2} (\partial_{x_1}^2 + \partial_{x_2}^2) + \frac{1}{2}
(\frac{ x_1^2}{\nu_1^2} + \frac{ x_2^2}{\nu_2^2}) \qquad, 
\end{equation}
where, $\partial_{x_i} \equiv \frac{\partial}{\partial x_i}$ for $i = 1, 2$
and $\nu_1^{-1}$ and $\nu_2^{-1}$ are the dimensionless frequencies such that
their ratio is a rational number. By a similarity transformation (ST), one
gets
\begin{equation} 
\tilde{H} \equiv \psi_0^{-1} H \psi_0 = \frac{1}{\nu_1} x_1 \partial_{x_1} +
\frac{1}{\nu_2} x_2 \partial_{x_2} - \frac{1}{2} \partial_{x_1}^2 - 
\frac{1}{2} \partial_{x_2}^2 + E_0 \qquad,
\end{equation}
where, $\psi_0 \equiv \exp{ (- \frac{1}{2 \nu_1} x_1^2 - \frac{1}{2 \nu_2}
x_2^2)}$ and $E_0 = \frac{1}{2 \nu_1} + \frac{1}{2 \nu_2}$. One more ST on
$\tilde{H}$ yields
\begin{equation} \label{bar}
\bar{H} \equiv \hat{T}^{-1} \tilde{H} \hat{T} = \frac{1}{\nu_1} x_1
\partial_{x_1} + \frac{1}{\nu_2} x_2 \partial_{x_2} + E_0 \qquad,
\end{equation}
here, $\hat{T} \equiv \exp{(- \frac{\nu_1}{4} \partial_{x_1}^2 -
\frac{\nu_2}{4} \partial_{x_2}^2)}$. 

By redefining $x_i = y_i^{\frac{1}{\nu_i}}$, (\ref{bar}) becomes,
\begin{equation} \label{bary}
\bar{H} = y_1 \partial_{ y_1} + y_2 \partial_{y_2} + E_0 \qquad;
\end{equation}
$\bar{H}$ can be easily connected to the isotropic oscillator by one more ST:
\begin{equation} \label{iso}
H_o \equiv \hat{U} \bar{H} \hat{U}^{-1} = - \frac{1}{2}
\partial_{ y_1}^2 - \frac{1}{2} \partial_{y_2}^2 + \frac{1}{2} y_1^2 +
\frac{1}{2} y_2^2  + E_0 - 1 \qquad,  
\end{equation}
where, $\hat{U} \equiv \exp(- \frac{1}{2} y_1^2 - \frac{1}{2} y_2^2)
\exp(- \frac{1}{4} \partial_{ y_1}^2 - \frac{1}{4} \partial_{ y_2}^2)$.

From (\ref{iso}), one can define $a_{o,i} = \frac{1}{\sqrt{2}} (y_i +
\partial_{y_i})$ and $a_{o,i}^\dagger = \frac{1}{\sqrt{2}} (y_i -
\partial_{y_i})$: 
$$H_o = \frac{1}{2} \{a_{o,1}\,\,,\,\,a_{o,1}^\dagger\} + \frac{1}{2}
\{a_{o,2}\,\,,\,\,a_{o,2}^\dagger\} + E_0 - 1 \qquad,$$  
with $[a_{o,i}\,\,,\,\,a_{o,j}^\dagger] = \delta_{ij}$ and
$[H_o\,\,,\,\,a_{o,i} (a_{o,i}^\dagger)] = - a_{o,i} (a_{o,i}^\dagger)$. By
an inverse ST, one can formally write down the spectrum generating algebra
for the original Hamiltonian in (\ref{aniso}) as
\begin{eqnarray}
H = \frac{1}{2} \{a_{y_1}^{-}\,\,,\,\,a_{y_1}^{+}\} + \frac{1}{2}
\{a_{y_2}^{-}\,\,,\,\,a_{y_2}^{+}\} + E_0 - 1
\qquad,
\end{eqnarray}
with 
\begin{eqnarray}\label{com1}
[a_{y_i}^{-}\,\,,\,\,a_{y_j}^{+}] = \delta_{ij}
\qquad,\nonumber
\end{eqnarray}
and
\begin{eqnarray}\label{com2}
[H\,\,,\,\,a_{y_i}^{\pm}] = \pm \,\,a_{y_i}^{\pm} \qquad. 
\end{eqnarray}
Here, $a_{y_i}^{-} = \hat{M} a_{o,i} \hat{M}^{-1}$, $a_{y_i}^{+} = \hat{M}
a_{o,i}^\dagger \hat{M}^{-1}$ and $\hat{M} \equiv \psi_0 \hat{T}
\hat{U}^{-1}$.

The complete set of eigenfunctions can be obtained
from (\ref{bar}) by using $a_{x_i}^{-} = (\psi_0 \hat{T}) \partial_{x_i}
(\psi_0 \hat{T})^{-1}$ and $a_{x_i}^{+} = (\psi_0 \hat{T}) x_i (\psi_0
\hat{T})^{-1}$. In this case, it can be verified that,
\begin{eqnarray}
&&H = \frac{1}{2 \nu_1} \{a_{x_1}^{-}\,\,,\,\,a_{x_1}^{+}\} + \frac{1}{2 \nu_2}
\{a_{x_2}^{-}\,\,,\,\,a_{x_2}^{+}\} \qquad, \nonumber\\
&&[H\,\,,\,\,a_{x_i}^{\pm}] = \pm \frac{1}{\nu_i} \,\,a_{x_i}^{\pm} \qquad.
\end{eqnarray}
The ground-state is obtained by $a_{x_i}^- \psi_0 = 0$ and the generic,
unnormalized, excited states are
\begin{equation}
\psi_{n_1,n_2} = (a_{x_1}^{+})^{n_1} (a_{x_2}^{+})^{n_2} \psi_0 \qquad,
\end{equation}
with energy eigenvalues 
\begin{equation} \label{n}
E_{n_1,n_2} = \frac{n_1}{\nu_1} + \frac{n_2}{\nu_2} + E_0 \qquad. 
\end{equation}
It is clear that degeneracy occurs only when $\frac{n_i}{\nu_i} = m_i; m_i =
0, 1, 2, \cdots$, in which case $E_{m_1,m_2} = m_1 + m_2 + E_0$.  In order to
identify the symmetry algebra underlying the degeneracy, it is convenient to
work in the $y$-basis. It is easy to see that $a_{y_i}^{-}
\psi_0 = 0$ and the degenerate excited states can be obtained by the repeated
application of $a_{y_i}^{+}$ on $\psi_0$:
\begin{equation}
\psi_{m_1,m_2} = (a_{y_1}^{+})^{m_1} (a_{y_2}^{+})^{m_2} \psi_0 \qquad,
\end{equation}
with energy eigenvalues 
\begin{equation}\label{m}
E_{m_1,m_2} = m_1 + m_2 + E_0 \qquad.
\end{equation}
The fact that the degenerate states are generated by the action of
$a_{y_i}^+$ on the ground-state and the Hamiltonian, in terms of $a_{y_i}^+$
and $a_{y_i}^-$ is precisely that of isotropic oscillators, makes the
identification of the symmetry behind the degeneracy completely transparent.
It is well known that the symmetry behind the planar isotropic oscillator is
$SU(2)$. One can now {\it unambiguously} construct an $SU(2)$ algebra
describing the degeneracy of the original anisotropic case:
\begin{equation}
J_{+} = a_{y_1}^{+} a_{y_2}^{-}\,,\,\,\,\,J_{-} = a_{y_2}^{+} a_{y_1}^{-}
\qquad, 
\end{equation}
and 
\begin{equation}
J_0 = \frac{1}{2} (a_{y_1}^{+} a_{y_1}^{-} - a_{y_2}^{+} a_{y_2}^{-})\qquad,
\end{equation}
such that
\begin{eqnarray}
{}[J_0\,\,,\,\,J_{\pm}] &=& \pm J_{\pm} \qquad, \nonumber\\
{}[J_{+}\,\,,\,\,J_{-}] &=& 2 J_0 \qquad.
\end{eqnarray}

At this moment, one is naturally led to the question of the construction of
the $SU(2)$ generators in terms of $a_{x_i}^+$ and $a_{x_i}^-$. It is easy to
see that the generators connecting the degenerate states can be written as
$$\tilde{J}_+ = (a_{x_1}^+)^{\nu_1}(a_{x_2}^-)^{\nu_2}\,\,\,,\mbox{and}\,\,\,
\tilde{J}_- = (a_{x_1}^-)^{\nu_1}(a_{x_2}^+)^{\nu_2}\qquad.$$ 
However, these generators do not lead to $SU(2)$ algebra, since
$[\tilde{J}_+\,\,,\,\,\tilde{J}_-]$ is not linear in $J_0$\cite{of,luc}.

In order to identify the regular $SU(2)$ generators, we prescribe below a
procedure which makes use of the method of canonical conjugate
(CC)\cite{ps,vs}. One can define the CC for $A_i \equiv
(a_{x_i}^{-})^{\nu_i}$, for $i = 1, 2$, as
\begin{equation}
[A_i\,\,,\,\,F_j^\dagger] = \delta_{ij} \qquad.
\end{equation}
Here, 
\begin{equation}
F_i^\dagger = \frac{1}{\nu_i} A_i^\dagger \frac{1}{A_i A_i^\dagger} (a_{x_i}^+
a_{x_i}^- + \nu_i) \qquad,
\end{equation}
and $A_i^\dagger \equiv (a_{x_i}^+)^{\nu_i}$. Using the above result, one can
define the $SU(2)$ generators,
\begin{equation}
\hat{J}^+ = F_1^\dagger A_2 \qquad,\qquad{}\hat{J}^- = F_2^\dagger A_1 \qquad,
\end{equation}
\begin{equation}
\hat{J}^0 = \frac{1}{2}(F_1^\dagger A_1 - F_2^\dagger A_2) = \frac{1}{2}
(A_1^\dagger A_1 - A_2^\dagger A_2) 
\end{equation}
such that
\begin{equation}
[\hat{J}_0\,\,,\,\,\hat{J}_{\pm}] = \pm \hat{J}_{\pm}\qquad,\qquad
[\hat{J}_+\,\,,\,\,\hat{J}_-] = 2 \hat{J}_0 \qquad.
\end{equation}
It can be easily checked that these generators commute with the Hamiltonian
(\ref{aniso}) and act properly in the degenerate space.

We now proceed to analyse the Hamiltonian of the most general two-dimensional
anisotropic oscillator with singular inverse-square terms:
\begin{equation} \label{saniso}
H_s = - \frac{1}{2} (\partial_{x_1}^2 + \partial_{x_2}^2) + \frac{1}{2}
(\frac{ x_1^2}{\nu_1^2} +
\frac{ x_2^2}{\nu_2^2}) + \frac{g_1^2}{2} x_1^{-2} + \frac{g_2^2}{2} x_2^{-2}
\qquad.
\end{equation}
Performing a ST by the ground-state wavefunction, $\psi_s$, one gets
\begin{equation}
\tilde{H}_s \equiv \psi_s^{-1} H \psi_s = \frac{1}{\nu_1} x_1 \partial_{x_1} +
\frac{1}{\nu_2} x_2 \partial_{x_2} - \hat{O_1} - \hat{O_2} + E_0 \qquad, 
\end{equation}
where, $\psi_s \equiv x_1^{\alpha_1} x_2^{\alpha_2} \exp{ (- \frac{1}{2
\nu_1} x_1^2 - \frac{1}{2 \nu_2} x_2^2)}$, $\alpha_i = \frac{1}{2} (1 + 
\sqrt{1 + 4 g_i^2})$, $\hat{O_i} \equiv \frac{1}{2} \partial_{x_i}^2 +
\frac{\alpha_i}{x_i} \partial_{x_i}$ and $E_0 \equiv \frac{1}{2 \nu_1} +
\frac{1}{2 \nu_2} + \frac{\alpha_1}{\nu_1} + \frac{\alpha_2}{\nu_2}$. One
more ST on $\tilde{H}$ yields
\begin{equation} \label{sbar}
\bar{H}_s \equiv \hat{T}_s^{-1} \tilde{H} \hat{T}_s = \frac{1}{\nu_1} x_1
\partial_{x_1} + \frac{1}{\nu_2} 
x_2 \partial_{x_2} + E_0 \qquad,
\end{equation}
here, $\hat{T}_s \equiv \exp{(- \frac{\nu_1}{2} \hat{O_1} - \frac{\nu_2}{2}
\hat{O_2})}$.

By redefining $x_i = y_i^{\frac{1}{2 \nu_i}}$, (\ref{sbar}) becomes,
\begin{equation}
\bar{H}_s = 2 (y_1 \partial_{ y_1} + y_2 \partial_{y_2}) + E_0 \qquad;
\end{equation}
$\bar{H}_s$ can be easily connected to the isotropic oscillator on a plane by
one more ST:
\begin{equation} \label{siso}
H_{\rm{iso}} \equiv \hat{U}_s \bar{H}_s \hat{U}_s^{-1} = - (\partial_{ y_1}^2 +
\partial_{y_2}^2) + y_1^2 + y_2^2  + E_0 - 2 \qquad,   
\end{equation}
where, $\hat{U}_s \equiv \exp(- \frac{1}{2} y_1^2 - \frac{1}{2} y_2^2)
\exp(- \frac{1}{4} \partial_{ y_1}^2 - \frac{1}{4} \partial_{ y_2}^2)$.

It is worth pointing out that the choice of the $y_i$ variables ensures
normalizability of the wavefunctions for $H_s$ when they are constructed from
those of the $H_{\rm iso}$.

In order to get the spectrum generating and degeneracy algebras, one can
proceed in parallel to the anisotropic oscillator without singular terms.
Defining $b_{x_i}^{-} [b_{y_i}^{-}] = (\psi_s \hat{T}_s) \partial_{x_i}
[\partial_{y_i}] (\psi_s \hat{T}_s)^{-1}$ and $b_{x_i}^{+} [b_{y_i}^{+}] =
(\psi_s \hat{T}_s) x_i [y_i](\psi_s \hat{T}_s)^{-1}$, it is easy to see that
\begin{eqnarray}\label{scom1}
[b_{x_i}^{-}\,\,,\,\,b_{x_j}^{+}] = [b_{y_i}^{-}\,\,,\,\,b_{y_i}^{+}] =
\delta_{ij} \qquad,
\end{eqnarray}
and
\begin{eqnarray}\label{scom2}
[H\,\,,\,\,b_{x_i}^{\pm} (b_{y_i}^{\pm})] = \pm \frac{1}{\nu_i}
\,\,b_{x_i}^{\pm} (\pm b_{y_i}^{\pm})\qquad. 
\end{eqnarray}

The ground-state wavefunction of the Hamiltonian $H_s$ can be gotten by
$b_{x_i}^{-} \psi_0 = b_{y_i}^{-} \psi_s = 0$ and all the excited states can
be obtained by the repeated application of $b_{x_i}^{+}$ on $\psi_s$.  The
generic, unnormalized, excited state can be written as
\begin{equation}
\Psi_{n_1,n_2} = (b_{x_1}^{+})^{2 n_1} (b_{x_2}^{+})^{2 n_2} \psi_s \qquad,
\end{equation}
and the corresponding energy eigenvalues are 
\begin{equation} \label{sn}
E_{n_1,n_2} = 2 (\frac{n_1}{\nu_1} + \frac{n_2}{\nu_2}) + E_0 \qquad. 
\end{equation}
As has been pointed out earlier, for the analysis of the degeneracy
structure, it is advantageous to work with the $b_{y_i}^{-}$ and
$b_{y_i}^{+}$ operators. The generic excited state obtained by the repeated
action of $b_{y_i}^+$ are
\begin{equation}
\psi_{m_1,m_2} = (b_{y_1}^{+})^{m_1} (b_{y_2}^{+})^{m_2} \psi_s \qquad,
\end{equation}
with energy eigenvalues 
\begin{equation}\label{sm}
E_{m_1,m_2} = 2 (m_1 + m_2) + E_0 \qquad.
\end{equation}
It is clear from (\ref{sn}) and (\ref{sm}) and also from (\ref{scom1}) and
(\ref{scom2}) that $b_{y_i}^{+}$ creates states whose eigenvalues are $\nu_i$
times the eigenvalues of the states created by $b_{x_i}^+$.

Using $b_{y_i}^{\pm}$, one can construct the generators of $SU(2)$ algebra in
a manner analogous to the previously treated example.

It is easy to see that the following generators
$$\tilde{J}_+ = (b_{x_1}^{+})^{2 \nu_1}(b_{x_2}^-)^{2
\nu_2}\,\,\,,\mbox{and}\,\,\, 
\tilde{J}_- = (b_{x_1}^-)^{2 \nu_1}(b_{x_2}^{+})^{2 \nu_2}\qquad,$$ 
do not belong to the $SU(2)$ algebra, since
$[\tilde{J}_+\,\,,\,\,\tilde{J}_-]$ is not linear in $J_0$\cite{of,luc}. 

Akin to the anisotropic oscillator without singular terms, the regular
$SU(2)$ generators can be constructed by the method of CC\cite{ps,vs}. One
can define the CC for $B_i \equiv (b_{x_i}^-)^{2
\nu_i}$, for $i = 1, 2$, as 
\begin{equation}
[B_i\,\,,\,\,G_j^\dagger] = \delta_{ij}
\end{equation}
Here, 
\begin{equation}
G_i^\dagger = \frac{1}{2 \nu_i} B_i^\dagger \frac{1}{B_i
B_i^\dagger} (b_{x_i}^+ b_{x_i}^- + 2 \nu_i) \qquad,
\end{equation}
and $B_i^\dagger \equiv (b_{x_i}^+)^{2 \nu_i}$. Using these $B_i$ and
$G_i^\dagger$, we can define the $SU(2)$ generators in exactly the same way
as that of the earlier example.
 
Here, we would like to remark that, when $\nu_1 = \nu_2 = 1$, $x_1
\rightarrow \frac{1}{\sqrt{2}} (x_1 - x_2)$ and $x_2 \rightarrow
\frac{1}{\sqrt{2}} (x_1 + x_2)$, the model in (\ref{saniso}) goes to a
special case of the two particle Calogero model of $B_N$ type\cite{guru2}.
This can be further reduced to the two particle Calogero model of $A_{N-1}$
type\cite{guru1} by choosing $g_2 = 0$. Hence, our analyses are also
applicable to the above CS type models; one needs only to symmetrize the
wavefunctions appropriately.

The above analysis for the two-dimensional anisotropic oscillator can be
immediately generalized to the $N$ dimensional case. The Hamiltonian with
singular terms reads as
\begin{equation}
H = - \frac{1}{2} \sum_{i=1}^N \partial_{x_i}^2 + \frac{1}{2} \sum_{i=1}^N
\frac{x_i^2}{\nu_i^2} + \sum_{i=1}^N \frac{g_i^2}{2} x_i^{-2} \qquad. 
\end{equation}
The following ST by $\hat{S} \equiv \prod_{i=1}^N x_i^{\alpha_i} \exp{ (-
\frac{1}{2} \sum_{i=1}^N \frac{x_i^2}{\nu_i})} \exp(- \sum_{i=1}^N
\frac{1}{2} \nu_i \hat{P_i})$; where, $\hat{P_i} \equiv \frac{1}{2}
\partial_{x_i}^2 + \frac{\alpha_i}{x_i} \partial_{x_i}$, yields 
\begin{equation}\label{nbarx}
\bar{H} \equiv \hat{S}^{-1} \tilde{H} \hat{S} = \sum_{i=1}^N \frac{1}{\nu_i}
x_i \partial_{x_i} + E_0 \qquad,  
\end{equation}
where, $\alpha_i = \frac{1}{2} (1 + \sqrt{1 + 4 g_i^2})$ and $E_0 \equiv
\sum_{i=1}^N (\frac{1}{2 \nu_i} + \frac{\alpha_i}{\nu_i})$. Akin 
to the two-dimensional case, one can redefine $x_i = y_i^{\frac{1}{2
\nu_i}}$: (\ref{nbarx}) becomes,
\begin{equation} \label{nbary}
\bar{H} = 2 \sum_{i=1}^N y_i \partial_{ y_i} + E_0 \qquad.
\end{equation}
Now, (\ref{nbary}) can be shown to be equivalent to an isotropic harmonic
oscillator in $N$ dimension, with equal frequencies by the following ST,
\begin{equation}
H_o \equiv \hat{U} \bar{H} \hat{U}^{-1} = - \sum_{i=1}^N \partial_{ y_i}^2 +
\sum_{i=1}^N y_i^2 + E_0 - N \qquad,  
\end{equation}
where, $\hat{U} \equiv \exp(- \frac{1}{2} \sum_i y_i^2) 
\exp(- \frac{1}{4} \sum_i \partial_{ y_i}^2)$. 

A procedure similar to the two-dimensional case can be straightforwardly
employed to show that the energy level degeneracy of the anisotropic
oscillator in $N$ dimensions is described by an $SU(N)$ algebra. The
generators, analogous to the Fradkin tensor\cite{frad} for the isotropic
case, are given by 
$$ F_{ij} = c_{y_i}^+ c_{y_j}^- \qquad,$$ 
where,
$c_{y_i}^- = \hat{S} \partial_{y_i} \hat{S}^{-1}$ and $c_{y_i}^+ =
\hat{S} y_i \hat{S}^{-1}$. Another basis for the Fradkin tensor can be given
by the method of canonical conjugate as 
$$F_{ij}^\prime = K_i^\dagger C_j \qquad,$$
where, 
$$C_i \equiv (c_{x_i}^-)^{2 \nu_i}\,\,\,,\,\,\,K_i^\dagger = \frac{1}{2 \nu_i}
C_i^\dagger \frac{1}{C_i C_i^\dagger} (c_{x_i}^+ c_{x_i}^- + 2 \nu_i)\qquad.$$
and
$$[C_i\,\,,\,\,K_j^\dagger] = \delta_{ij}\qquad.$$
Here, $C_i^\dagger \equiv (c_{x_i}^+)^{2 \nu_i}$, $c_{x_i}^- = \hat{S}
\partial_{x_i} \hat{S}^{-1}$ and $c_{x_i}^+ = \hat{S} x_i \hat{S}^{-1}$.

The above analysis can be easily applied to the anisotropic oscillator with
commensurate frequencies and without singular inverse-square terms.

\newpage
\begin{center}
{\bf 3. MANY-BODY INTERACTING MODELS}
\end{center}

In the following, we first analyse a recently studied one-dimensional model
of four identical particles with both two-body and four-body inverse-square
interactions given by the Hamiltonian\cite{ruhl},
\begin{equation}
H = - \frac{1}{2} \sum_{i=1}^4 \partial_{x_i}^2 + \frac{1}{2} \sum_{i=1}^4
x_i^2 + g_1 \sum_{{i,j}\atop {i\ne j}} (x_i - x_j)^{-2} + g_2 \sum_{{3 {\rm
independent}}\atop {{\rm terms}}} (x_i + x_j - x_k - x_l)^{-2} \qquad.
\end{equation}
The correlated bosonic ground-state of $H$ is given by
$$\psi_0 = \prod_{i<j} |x_i - x_j|^\alpha
\,\,\,\prod_{\rm 3 indep. terms} (x_i + x_j - x_k - x_l)^\beta
\,\,\,\exp\{- \frac{1}{2} \sum_i x_i^2\}\qquad,$$
where, $\alpha = \frac{1}{2} (1 + \sqrt{1 + 4 g_1})$ and $\beta = \frac{1}{2}
(- 1 + \sqrt{1 + 2 g_2})$. By performing a ST on $H$ with respect to
$\psi_0$, one gets
\begin{equation}
\psi_0^{-1} H \psi_0 \equiv \tilde{H} = \sum_i x_i \partial_{x_i} - \hat{A} +
E_0 \qquad, 
\end{equation}
where,
$$\hat{A} = \frac{1}{2} \sum_i \partial_{x_i}^2 + \alpha \sum_{i\ne j}
\frac{1}{x_i - x_j} \partial_{x_i} + \beta \sum_{3 {\rm indep. terms}}
\frac{1}{(x_i + x_j - x_k - x_l)} (\partial_{x_i} + \partial_{x_j} -
\partial_{x_k} - \partial_{x_l})$$
and $E_0 = 2 + 6 \alpha + 3 \beta$. Another ST by $\hat{S} = \exp\{-
\hat{A}/2\}$ on $\tilde{H}$ diagonalizes it completely:
\begin{equation} \label{eul}
\hat{S}^{-1} \tilde{H} \hat{S} \equiv \bar{H} = \sum_i x_i \partial_{x_i} +
E_0 \qquad.
\end{equation}
The explicit connection of $H$ with the decoupled oscillators can be obtained
by one more ST on $\bar{H}$
\begin{equation}
\hat{T}^{-1} \bar{H} \hat{T} = - \frac{1}{2} \sum_{i=1}^4 \partial_{x_i}^2 +
\frac{1}{2} \sum_{i=1}^4 x_i^2 + E_0 - 2 \qquad,
\end{equation}
where, $\hat{T} \equiv \exp(- \frac{1}{2} \sum_{i=1}^4 x_i^2)
\exp(- \frac{1}{4} \sum_{i=1}^4 \partial_{ x_i}^2)$.

Akin to the anisotropic oscillator case, one can find the creation and
annihilation operators for the above four-particle model:
\begin{equation}
H = \sum_{i=1}^4 H_i + E_0 - 2 =
\frac{1}{2} \sum_{i=1}^4 \{a_i^-\,\,,\,\,a_i^+\} + E_0 - 2 \qquad,
\end{equation}
where, $H_i = \frac{1}{2} \{a_i^-\,\,,\,\,a_i^+\}$, $a_i^{-} = \hat{M} b_i
\hat{M}^{-1}$, $a_i^{+} = \hat{M} b_i^\dagger \hat{M}^{-1}$, $b_i =
\frac{1}{\sqrt{2}} (x_i + \partial_{x_i})$, $b_i^\dagger = \frac{1}{\sqrt{2}}
(x_i - \partial_{x_i})$ and $\hat{M}
\equiv \psi_0 \hat{S} \hat{T}^{-1}$. It follows that
$$[a_i^{-}\,\,,\,\,a_j^{+}] = \delta_{ij} \qquad,$$ 
and 
$$[H_i\,\,,\,\,a_i^{-} (a_i^{+})] = [H\,\,,\,\,a_i^{-} (a_i^{+})] = -
a_i^{-} (a_i^{+}) \qquad.$$ 

For the construction of the eigenfunctions of this model, one can also make
use of (\ref{eul}) since, $x_i$ and $\partial_{x_i}$ serve as the creation
and annihilation operators respectively. The ground-state can be chosen:
$\partial_{x_i} \phi_0 = 0 \,\,,\,\,\mbox{for}\,\, i=1,2,3,4$. The excited
states are given by the monomials $\prod_i^4 x_i^{n_i}$; $n_i = 0, 1,
2,\cdots$. It is worth mentioning that for the normalizability of the
wavefunctions, one needs to check that the action of $\exp\{- \hat A/2\}$ on
an appropriate symmetric combination of the eigenstates of $\sum_{i=1}^4 x_i
\partial_{x_i}$ yields a polynomial solution\cite{guru1}. In the following,
we present explicit forms of some eigenstates.

\noindent {\it Case I}: Eigenstates corresponding to the center-of-mass
degree of freedom: 
\begin{equation}
\psi_{n_1,0,0,0} = \psi_0\,\,\exp\{- \frac{1}{2} \hat{A}\}
R^{n_1} = \exp\{- \frac{1}{4} \sum_{N=1}^4 \partial_{x_i}^2\}
\,\,R^{n_1}\qquad, 
\end{equation}
where, $R = 4^{-1} (x_1 + x_2 + x_3 + x_4)$.
This can be cast in the form\cite{kok},
\begin{equation}
\psi_{n_1,0,0,0} = 8^{- n_1} n_1! \psi_0\,\,\sum_{\sum_{i=1}^4 m_i = n_1}
\prod_{i=1}^4 \frac{H_{m_i}(x_i)}{m_i!} \qquad.
\end{equation}
Here, $H_{m_i}(x_i)$ are the Hermite polynomials.

\noindent {\it Case II}: Eigenstates corresponding to the radial degree of
freedom\cite{ak}: 
\begin{equation}
\psi_{0,n_2,0,0} = \psi_0\,\,\exp\{- \frac{1}{2} \hat{A}\}
(r^2)^{n_2} \qquad.
\end{equation}
where, $r^2 \equiv \frac{1}{4} \sum_{{i,j = 1}\atop {i < j}}^4 (x_i -
x_j)^2$.  It is easy to check that $\hat{A} (r^2)^n = 2 n (n + \frac{1}{2} +
6 \alpha + 3 \beta) (r^2)^{n - 1}$; using this, one gets
\begin{eqnarray}
\psi_{0,n_2,0,0} &=& \psi_0\,\,(-1)^{n_2}\,\,n_2!\,\,\sum_{l=0}^{n_2}
\frac{(-1)^{n_2}}{l! (n_2 - l)!} \frac{(n_2 + \frac{1}{2} + 6 \alpha + 3
\beta)!}{(l + \frac{1}{2} + 6 \alpha + 3 \beta)!} (r^2)^l \nonumber\\
&=& \psi_0\,\,(-1)^{n_2}\,\,n_2!\,\,L_{n_2}^{\frac{1}{2} + 6 \alpha + 3
\beta}(r^2)\qquad. 
\end{eqnarray}
Here, $L_{n_2}^{\frac{1}{2} + 6 \alpha + 3 \beta}(r^2)$ is the Lagurre
polynomial. In a similar way, one can find any generic excited state
wavefunction.

One can also define $<<0|{S}_n(\{a_i^-\}) = <<n|$ and ${S}_n(\{a_i^+\})|0> =
|n>$ as the bra and ket vectors; ${S}_n$ is a symmetric and homogeneous
function of degree $n$ and $<<0|a_i^+ = a_i^- |0> = 0$. Since the oscillators
are decoupled, the inner product between these bra and ket vectors proves
that any ket $|n>$, with a given partition of $n$, is orthogonal to all the
bra vectors, with different $n$ and also to those with different partitions
of the same $n$.

The quantum integrability of this model can be proved easily by noting that
the set $\{H_1, H_2, H_3, H_4\}$ provides the four mutually commuting
conserved quantities, {\i.e.}, $[H\,\,,\,\,H_k] = [H_i\,\,,\,\,H_j] = 0$.
From this set, one can construct four linearly independent symmetric
conserved quantities.

More general interacting models of the CSM type, which can be mapped to
decoupled oscillators, can be constructed in the following manner.

One starts with the general Hamiltonian of the type
\begin{equation}\label{gh}
H = - \frac{1}{2} \sum_{i=1}^N \partial_{x_i}^2 + V(x_1,x_2,x_3,\cdots,x_N)
\qquad, 
\end{equation}
having $\Phi_0$ as the ground-state wavefunction and 
$V(x_i,x_2,x_3,\cdots,x_N) = \frac{1}{2 \Phi_0} \sum_{i=1}^N \partial_{x_i}^2
\Phi_0$. In order to bring $H$ to the following form
\begin{equation} \label{gtilde}
\Phi_0^{-1} H \Phi_0 \equiv \tilde{H} = \sum_i^N x_i \partial_{x_i} - \hat{A}
+ E_0 \qquad, 
\end{equation}
the ground-state wavefunction must be of the form $\Phi_0 = G J$; where, $G
\equiv \exp\{- \frac{1}{2} \sum_{i=1}^N x_i^2\}$,
$\hat{A} \equiv \frac{1}{2} \sum_{i=1}^N \partial_{x_i}^2 + \frac{1}{J}
\sum_{i=1}^N \partial_{x_i}(\ln J) \partial_{x_i}$ and $E_0$ is the
ground-state energy. 

$\tilde{H}$ can be mapped to the Euler operator by another ST
\begin{equation} 
\hat{S}^{-1} \tilde{H} \hat{S} \equiv \bar{H} = \sum_{i=1}^N x_i
\partial_{x_i} + E_0 \qquad,  
\end{equation}
provided, the following equation holds,
\begin{equation}
\left[\tilde H , \exp\{- \hat A/2\}\right] = 
\left[\sum_i x_i {\partial}_{x_i} , \exp\{- \hat A/2\}\right] 
= \hat A \exp\{- \hat A/2\} \qquad.
\end{equation}
The above condition restricts $J$ to be a homogeneous function of the
particle coordinates.

Now, it is easy to see that, the Hamiltonian in (\ref{gh}) can be mapped to
free oscillators by a series of STs,
\begin{equation}
G \hat{E} \exp\{\hat A/2\} {\Phi_0}^{-1} H \Phi_0 \exp\{- \hat A/2\} {\hat
E}^{-1} G^{-1} = -\frac{1}{2} \sum_i \partial_{x_i}^2 +
\frac{1}{2} \sum_i x_i^2 + (E_0 - \frac{1}{2} N) \qquad,
\end{equation}
where,  $\hat E \equiv \exp\{-\frac{1}{4} \sum_{i=1}^N \partial_{x_i}^2\}$.

However, it is important to check that, the action of $\exp\{- \hat A/2\}$ on
an appropriate linear combination of the eigenstates of $\sum_{i=1}^N x_i
\partial_{x_i}$ yields a polynomial solution. Otherwise, the
resulting functions are not normalizable with respect to $\Phi_0^2$ as a
weight function. Appropriate choices of $J$ will yield new solvable models
having linear spectra. This case can also be generalized to the
higher-dimensional interacting models\cite{guru1,sr,pkg}.

\begin{center}
{\bf 4. LAUGHLIN WAVEFUNCTION AND DECOUPLED HARMONIC OSCILLATORS}
\end{center}

In this section, we study some planar many-body Hamiltonians relevant for the
description of the quantum Hall effect. These Hamiltonians describe electrons
in a magnetic field, with two-body and three-body inverse-square interactions
arising due to Chern-Simons gauge field and have Laughlin
wavefunction\cite{lin} as the ground-state\cite{srg}. We explicitly prove
that, this model can be exactly mapped to a set of free harmonic oscillators
on the plane. As a consequence, the existence of a {\it linear}
$W_{1+\infty}$ algebra with Laughlin wavefunction as its highest weight
vector, is pointed out in a rather elegant and straightforward manner.

The relevant Hamiltonian is given by
\begin{eqnarray} \label{iz}
H =&& \frac{1}{2} \sum_{i=1}^N ( - 4 \partial_{{\bar{z}}_i} \partial_{z_i} +
z_i \partial_{z_i} - \bar{z}_i \partial_{\bar{z}_i} + \frac{1}{4} \bar{z}_i
z_i ) + 2 \eta \sum_{{i=1\atop {i\ne j}}}^N \frac{1}{(z_i - z_j)}
(\partial_{\bar{z}_i} - \frac{1}{4} z_i) \nonumber\\
&-& 2 \eta \sum_{{i=1\atop {i\ne j}}}^N \frac{1}{(\bar{z_i} - \bar{z_j})}
(\partial_{z_i} - \frac{1}{4} \bar{z_i})
+ 2 \eta^2 \sum_{{i,j,k=1}\atop
{i\ne j ; i\ne k}}^N \frac{1}{(z_i - z_j) (\bar{z}_i - \bar{z}_k)}
\qquad, 
\end{eqnarray}
where, $z_i = x_i + i y_i$.
The ground-state of this model was found to be of Laughlin form,
$$\psi_0 =
\prod_{i<j} (z_i - z_j)^\eta \,\,\exp\{- \frac{1}{4} \sum_i \bar{z}_i
z_i\}\qquad.$$ 
By performing a ST, one gets
\begin{eqnarray} \label{st}
\psi_0^{-1} H \psi_0 \equiv \tilde{H} = \sum_i z_i \partial_{z_i} - \hat{A}
+ \frac{1}{2} N \qquad,
\end{eqnarray}
where, $\hat{A} \equiv 2 \sum_i \partial_{{\bar{z}}_i} 
\partial_{z_i} + 2 \eta \sum_{i\ne j} (\frac{1}{\bar{z_i} - \bar{z_j}}
\partial_{z_i}- \frac{1}{z_i - z_j}
\partial_{{\bar{z}}_i})$.
It is easy to check that
\begin{equation}
[\sum_i z_i \partial_{z_i}\,\,,\,\,\hat{A}] = - \hat{A} + 4 \pi \eta
\sum_{{i,j}\atop {i\ne j}} (z_i - z_j) \delta^2(z_i - z_j) \partial_{z_i}
\qquad. 
\end{equation}
In the view of the $\delta$-function identity {\it i.e.}, $x \delta(x) = 0$,
the above equation reduces to
\begin{equation}
[\sum_i z_i \partial_{z_i}\,\,,\,\,\hat{A}] = - \hat{A} \qquad. 
\end{equation}
Performing a ST by $\exp\{- \hat{A}\}$, (\ref{st}) becomes
\begin{eqnarray}\label{fz}
\exp\{\hat{A}\}\,\,\tilde{H}\,\,\exp\{- \hat{A}\} \equiv \bar{H} = \sum_i z_i
\partial_{z_i} + \frac{1}{2} N \qquad.
\end{eqnarray}
Finally, the following ST by $\hat{W} \equiv \exp\{2 \sum_i
\partial_{{\bar{z}}_i} \partial_{z_i}\} \exp\{\frac{1}{4} \sum_i \bar{z}_i
z_i\}$ brings the above Hamiltonian to a Hamiltonian of $N$ free
harmonic oscillators,
\begin{equation}
\hat{W}^{-1} H \hat{W} = \frac{1}{2} \sum_{i=1}^N ( - 4
\partial_{{\bar{z}}_i} \partial_{z_i} + z_i \partial_{z_i} - \bar{z}_i
\partial_{\bar{z}_i} + \frac{1}{4} \bar{z}_i z_i) \qquad.
\end{equation}
By defining $a_i^+ = {\hat{S}}^{-1} z_i \hat{S}$ and $a_i^- = {\hat{S}}^{-1}
\partial_{z_i} \hat{S}$; where $\hat{S} = \psi_0 \exp\{\hat{A}\}$, and making
use of (\ref{fz}), one can rewrite 
(\ref{iz}) as 
\begin{equation}
H = \sum_i H_i + \frac{1}{2} N = \sum_i a_i^+ a_i^- + \frac{1}{2} N
\end{equation}
where, $H_i \equiv a_i^+ a_i^-$, such that
$$[a_i^-\,\,,\,\,a_j^+] = \delta_{ij}\qquad,$$
and 
$$[H_i\,\,,\,\,a_j^{\pm}] = \pm a_j^\pm \delta_{ij}\qquad.$$
These $N$ quantities $H_i$ serve as the conserved quantities which are in
involution, {\it i.e.}, $[H_i\,\,,\,\,H_j] = 0$. 

Since $a_i^-$ and $a_i^+$ obey non-interacting oscillator algebra, one can
make use of this fact to define a {\it linear} $W_{1+\infty}$ algebra.  The
generators of the $W_{1+\infty}$ algebra, $L_{m,n} =
\sum_{i=1}^N {(a_i^+)}^{m+1} {(a_i^-)}^{n+1}$, for $m, n \ge - 1$ obey the
linear relation\cite{cap}
\begin{equation}
[L_{m,n}\,\,,\,\,L_{r,s}] = \sum_{p=0}^{\mbox{Min}(n,r)} \frac{(n+1)!
(r+1)!} {(n-p)! (r-p)! (p+1)!} L_{m+r-p,n+s-p} - (n \leftrightarrow s, m
\leftrightarrow r) \qquad.
\end{equation}
The highest weight vector obtained from $L_{m,n} \psi_0 = 0$ for $n > m
\ge -1$ is nothing but the Laughlin wavefunction. 

In the following, we present two more two-dimensional models which can also
be made equivalent to a set of decoupled oscillators. The first one is due to
Rajaraman and Sondhi (RS)\cite{son}
\begin{eqnarray}
H_{RS} =&& \frac{1}{2} \sum_{i=1}^N ( - 4 \partial_{{\bar{z}}_i} \partial_{z_i} +
z_i \partial_{z_i} - \bar{z}_i \partial_{\bar{z}_i} + \frac{1}{4} \bar{z}_i
z_i ) + 2 \eta \sum_{{i=1\atop {i\ne j}}}^N \frac{1}{(z_i - z_j)}
(\partial_{\bar{z}_i} - \frac{1}{4} z_i) \qquad,
\end{eqnarray}
with the Laughlin wavefunction as the ground-state,
$$\psi_0 =
\prod_{i<j} (z_i - z_j)^\eta \,\,\exp\{- \frac{1}{4} \sum_i \bar{z}_i
z_i\}\qquad.$$ 
The second one can be viewed as a special case of (\ref{iz}):
\begin{eqnarray}
H =&& \frac{1}{2} \sum_{i=1}^N ( - 4 \partial_{{\bar{z}}_i} \partial_{z_i} +
z_i \partial_{z_i} - \bar{z}_i \partial_{\bar{z}_i} + \frac{1}{4} \bar{z}_i
z_i ) + 2 \eta \sum_{{i=1\atop {i\ne j}}}^N \frac{1}{(z_i - z_j)}
(\partial_{\bar{z}_i} - \frac{1}{4} z_i) \nonumber\\
&+& 2 \eta^2 \sum_{{i,j,k=1}\atop
{i\ne j ; i\ne k}}^N \frac{1}{(z_i - z_j) (\bar{z}_i - \bar{z}_k)}
\qquad, 
\end{eqnarray}
with the complex conjugate of the Laughlin wavefunction as the ground-state,
$$\psi_0 = \prod_{i<j} (\bar{z}_i - \bar{z}_j)^\eta \,\,\exp\{- \frac{1}{4}
\sum_i \bar{z}_i z_i\}\qquad.$$ 

In parallel to the earlier treatment, the above two Hamiltonians can be
mapped to a set of $N$ decoupled harmonic oscillators on a plane.  One can
also show the existence of a {\it linear} $W_{1+\infty}$ algebra for these
models.

\begin{center}
{\bf 5. CONCLUSION}
\end{center}

In conclusion, we have studied the degeneracy structure of the anisotropic
oscillator with rational frequency ratio, both with and without singular
terms, by mapping the given anisotropic oscillator to a regular isotropic
harmonic oscillator with equal frequencies.  This correspondence enabled one
to easily write down the $SU(N)$ symmetry algebra for the $N$-dimensional
anisotropic oscillator. This was done in two different ways and the $SU(N)$
generators are free from the ambiguities encountered earlier.

A similar technique was also applied to a recently studied four-particle
model in one dimension, with both pair-wise and four-body inverse-square
interactions and it was shown that the underlying algebraic structure of this
model is nothing but that of the free harmonic oscillators. We explicitly
proved the quantum integrability of this model and also computed some
eigenfunctions in the Cartesian basis. A method to construct new solvable
models was also outlined. It will be of great interest to study these one and
higher dimensional models explicitly. We further studied two-dimensional
models having Laughlin type wavefunctions as ground-state and establish its
connection with free oscillators. This enabled us to realize the Laughlin
wavefunctions as the highest weight vectors of a linear $W_{1+\infty}$
algebra. This analysis also needs further study since some of these models
can be used to implement the Jain picture of the fractional quantum Hall
effect. We hope to come back to some of these points in future work.

\begin{center}
{\bf 6. ACKNOWLEDGEMENTS}
\end{center}

The authors would like to acknowledge useful discussions with Prof. V.
Srinivasan. We would like to thank Prof. A. Khare for valuable suggestions
and also for bringing some of the references to our attention. N.G thanks
U.G.C (India) for financial support through the S.R.F scheme.

\end{document}